\UseRawInputEncoding
\documentclass[twocolumn]{aastex63}
\usepackage{mathptmx}

\newcommand{\teff}{T_{\rm eff}}

\newcommand{\vsini}{v\sin{i}}
\newcommand{\sini}{\sin{i}}

\mathchardef\mhyphen="2D

\graphicspath{{./}{figures/}}

\received{26 May 2021 }
\revised{5 July 2021}
\accepted{12 July 2021}

\shorttitle{Evidence for Tidal Spin-Up}
\shortauthors{Tejada Arevalo, Winn, Anderson}

\begin{document}

\title{{\bf Further Evidence for Tidal Spin-Up of Hot Jupiter Host Stars}}

\correspondingauthor{Roberto A. Tejada Arevalo}
\email{arevalo@princeton.edu}

\author[0000-0001-6708-3427]{Roberto A.\ Tejada Arevalo}
\affiliation{Department of Astrophysical Sciences, Princeton University, 4 Ivy Lane, Princeton, NJ 08544, USA}

\author[0000-0002-4265-047X]{Joshua N.\ Winn}
\affiliation{Department of Astrophysical Sciences, Princeton University, 4 Ivy Lane, Princeton, NJ 08544, USA}

\author[0000-0002-7388-0163]{Kassandra R.\ Anderson}
\altaffiliation{Lyman Spitzer, Jr.\ Postdoctoral Fellow}
\affiliation{Department of Astrophysical Sciences, Princeton University, 4 Ivy Lane, Princeton, NJ 08544, USA}

\begin{abstract}

For most hot Jupiters around main-sequence Sun-like stars,
tidal torques are expected to transfer angular
momentum from the planet's orbit to the star's rotation.
The timescale for this process is difficult to calculate,
leading to uncertainties in the history of orbital
evolution of hot Jupiters.
We present evidence for tidal spin-up by
taking advantage of recent advances in planet detection
and host-star characterization.
We compared the projected rotation
velocities and rotation periods
of Sun-like stars with hot Jupiters and spectroscopically similar stars with
(i) wider-orbiting 
giant planets, and (ii) less massive planets.
The hot Jupiter hosts tend to spin faster than the stars in either of the control samples.
Reinforcing earlier studies, the results imply that hot Jupiters alter
the spins of their host stars while they are on the main sequence, and
that the ages of hot-Jupiter hosts cannot be reliably
determined using gyrochronology.

\end{abstract}

\keywords{planetary systems, stars -- exoplanets, rotation, dynamical evolution and stability, gaseous planets, stellar ages, Hot Jupiters, tidal interactions}

\section{Introduction}
\label{sec:intro}

Dissipative tidal interactions between the two stars in a
close binary tend to align the stars' rotation axes,
circularize their orbit, and synchronize their spins with the orbit \citep[e.g.,][]{Zahn1977,Hut1981,Ogilvie2014}.
The evidence for these processes, as reviewed by \citet{Mazeh2008},
is based on measurements of the orbital and rotational properties
of binaries of various ages and evolutionary states.
Given the evidence for these tidal effects in stellar binaries,
we expect similar interactions to occur between
close-orbiting planets and their host stars.
The effects should be strongest for planets with
relatively high masses and small orbital separations: hot Jupiters.

Soon after the discovery of 51\,Pegasi\,b, \cite{Rasio1996} drew attention to the importance of tidal effects for hot Jupiters. The planet's low orbital
eccentricity was naturally explained as a consequence of tidal dissipation
within the planet's interior. Subsequent observations of hundreds
of hot Jupiters around main-sequence FGK stars
have confirmed that the orbital eccentricity tends
to be low when the orbital period is shorter than 10 days
\citep[see, e.g., Figure 3 of][]{Winn2015}.

Even after circularization, tidal torques should
continue to transfer angular momentum from the orbit
to the star due to tides on the star by the planet,
shrinking the orbit and spinning up the star.
However, for 51\,Peg\,b and many other hot Jupiters,
the reservoir of orbital angular momentum is too small to synchronize the
star.  Instead, the planet should spiral inward
and be destroyed when it crosses the Roche radius \citep{Levrard2009,Matsumura2010}.
The rate of this process is poorly known
because of uncertainties in the physics of tidal
dissipation as well as the star's steady loss of angular momentum
due to the magnetized stellar wind \citep[i.e.,``magnetic braking'';][]{WitteSavonjie2002,Barker2009,Ferraz-Mello2015a,Damiani2015}.

There are a few special cases of direct
observational evidence for tidal spin-up and orbital decay.
For example, the $\tau$\,Boo system
appears to be synchronized \citep{Butler1997},
and the orbital period of WASP-12b is shrinking
\citep{Maciejewski2016,Yee2019}.
There is also population-level 
evidence for tidal interactions.
For example, \cite{Jackson2009} and \cite{Cameron2018}
found that invoking tidal decay helped to
explain the observed distribution of orbital separations
of a large sample of hot Jupiters.
Likewise, \cite{Penev2018} reproduced the observed orbital
and rotational properties of a sample of 188 hot Jupiters
using a model for secular tidal evolution.

Our work was motivated by the desire to seek
less model-dependent evidence for tidal spin-up hot Jupiter host stars,
building on earlier work by \cite{Brown2014} and \cite{Maxted2015}.
Those authors framed the problem as a comparison between
the results of two methods for estimating a star's main-sequence age:
fitting the observable properties to the outputs of stellar-evolutionary models
(the ``isochrone age''), and assuming the star's rotation rate has slowed
down over time in the usual manner
(the ``gyrochronological'' or ``gyro'' age).
\cite{Brown2014} examined a sample of 68 hot Jupiter hosts and
found a
tendency for the gyro ages to be younger than the isochrone ages.
\cite{Maxted2015} obtained
a similar result with a sample of 28 hot Jupiters.
However, in neither case could the authors establish
a correlation between the size of the age discrepancy
and the mass ratio or orbital separation, the parameters
that should strongly influence the tidal dissipation rate.
These studies also left open the possibility that the discrepancy
between isochrone and gyro ages
reflected systematic errors or limitations in the cross-calibration
of these methods,
rather than a physical effect. 
 
Recent developments allowed us to improve on these earlier studies
by using a larger sample of planets,
constructing large samples of ``control stars'' without hot Jupiters,
and taking advantage of \cite{GaiaCollaboration2018} data for homogeneous determinations
of the basic stellar properties.
Section~\ref{sec:methods} describes the construction of our samples
of stars with hot Jupiters as well as control samples consisting of stars with wider-orbiting
or smaller planets. 
Section~3 presents the comparison of the projected rotation
velocities and rotation periods of the stars in the samples,
highlighting the evidence for faster rotation among the hot Jupiter hosts.
Section~\ref{sec:discussion} summarizes these results
and discusses implications for our understanding of tidal dissipation
and of hot Jupiters.

\section{Samples of Stars}
\label{sec:methods}

To simplify the interpretation of the
results, we focused on Sun-like stars,
defined here as main-sequence stars with effective temperatures
in the range from 5500 to 6000\,K.
At first, we imposed only one other criterion:
a limit on the surface gravity
of $\log g \geq 3.90$, to exclude evolved stars.
After an initial round of sample selection and analysis,
we imposed additional constraints on the surface gravity
($4.85 \geq \log g \geq 3.90$)
and metallicity ($0.44 \geq {\rm [Fe/H]} \geq -0.33$)
in order to ensure that the stars in our samples
had similar distributions of those parameters
(see Section \ref{subsec:spec_params}). For simplicity
of presentation, below we will describe only the samples
that resulted from these more restrictive criteria.

We wanted to examine
stars for which tidal spin-up is expected, as well as
stars for which it is not expected, and compare the observed
rotation properties.
To this end, we constructed a
sample of giant-planet host stars (as explained in
Section~\ref{subsec:giants}). Some of the giants are hot Jupiters,
while others are more distant giants for which tides are expected to be negligible.
We also constructed a sample of stars with
lower-mass planets for which tides are expected to be negligible (Section~\ref{subsec:smaller}).
We determined the masses, sizes, and ages
of all the stars in a homogeneous fashion (Section~\ref{subsec:ages}).
We defined a metric by which to rank the stars according to
the expected degree of tidal spin-up (Section~\ref{subsec:tidal_rank}).
We also made sure that the spectroscopic parameters and derived physical parameters of all
the stars in our samples span similar ranges,
to ensure that fair comparisons could be made (Section~\ref{subsec:spec_params}).

\subsection{Stars with Giant Planets}
\label{subsec:giants}

We began by merging the spectroscopic parameters
of the relatively homogeneous SWEET-Cat
catalog \citep[][]{Santos2013} with the more
comprehensive database of the NASA Exoplanet Archive \citep[NEA;][]{Akerson2013}\footnote{\url{https://exoplanetarchive.ipac.caltech.edu}} as of March 2021.
We selected the 273 stars satisfying our
effective temperature criteria from SWEET-Cat for which
the NEA reported at least one planet with a
mass exceeding 0.3 Jupiter masses. We discarded systems for which we did not find published $\vsini$ measurements, and kept 240 of these which satisfied further spectroscopic criteria (see Section~\ref{subsec:spec_params}).
About half of them are transiting planets, and the other half are
Doppler planets that are not known to transit.
We searched SWEET-Cat and the literature for all available information
about the projected rotation velocity ($v\sin i$) and
the rotation period
($P_{\rm rot}$) of these stars.


\subsection{Stars with smaller planets}
\label{subsec:smaller}

To construct a large sample of stars with low-mass planets,
we relied on the results of the California Kepler Survey \citep[CKS;][]{Petigura2017}.
We applied the spectroscopic criteria stated at the beginning
of this section, and required
all of the known planets to be smaller than 4 times the radius of Earth.
This resulted in a sample of 285 planets.

\subsection{Isochrone ages}
\label{subsec:ages}

The expected spin rate of a Sun-like star in the absence of tides
depends on age, due to the gradual effect of magnetic braking.
Therefore, in order to assess a star for any excess rotation,
we needed to know the stellar age. The ages of Sun-like
stars are famously difficult to determine because
their observable properties change little during the main-sequence
phase of stellar evolution.  For our samples, the only available
method for age determination is fitting the observable properties
to the outputs of stellar-evolutionary models (isochrone fitting)
which is subject to systematic errors due
to different choices and approximations in the
models and different choices of the observed quantities to match
with the models.  For maximum homogeneity,
we determined the isochrone
ages of all the stars in our samples with the same procedure. 
This approach allows for the most meaningful comparisons
between the calculated ages within our sample, even if the absolute
ages are still subject the usual systematic uncertainties of stellar
evolutionary modeling.

We used the \texttt{Isochrones} software package \citep{Morton2015},
which is based on the MESA Isochrones \& Stellar Tracks \citep{Choi2016}. We followed a
similar procedure as that described in Appendix A of \cite{Anderson2021}. 
We required the evolutionary models to match
the observed spectroscopic parameters
$\teff$, $\log{g}$ and [Fe/H] from SWEET-Cat, as well as the parallax and apparent magnitudes
($G$, $RP$, and $BP$) from \textit{Gaia} DR2.\footnote{\cite{Anderson2021}
used broadband photometry from 2MASS, \textit{WISE}, and \textit{Gaia} Data Release 2. We chose to fit only the \textit{Gaia} photometry.}
To avoid over-weighting any single input,
we adopted minimum uncertainties of $100$\,K in $\teff$, 0.1\,dex in [Fe/H],
0.1 mas in the parallax, and 0.01 in the apparent magnitudes.
We also excluded any apparent magnitudes with reported
uncertainties exceeding 0.1 mag. We enforced a
prior constraint on the 
extinction based on the value
obtained from the Galactic dust map MWDUST \citep{Bovy2015a}
for the star's coordinates and distance.
Table~\ref{tbl:data_t} gives a sample of the results for the stellar age, mass, and radius.

The reported uncertainties do not include the additional
systematic uncertainties inherent in the stellar-evolutionary models,
which are probably at least 10\%. Also, in this step we omitted the unusually young CoRoT-2 system, for which our isochrone analysis
disagreed strongly with previous results.

\begin{deluxetable*}{ccccccccccc}
\tabletypesize{\footnotesize}
\tablecaption{System Samples
\label{tbl:data_t}}
\tablecolumns{11}
\tablehead{
\colhead{Name} &
\colhead{$\vsini$ [km\,s$^{-1}$]} &
\colhead{$P_{\rm rot}$ [d]} &
\colhead{$\teff$ [K]} &
\colhead{Metallicity} &
\colhead{$\log{g}$} &
\colhead{Age [Gyr]} &
\colhead{$R$ [R$_\odot$]} &
\colhead{$M$ [M$_\odot$]} &
\colhead{$\log{\eta}$} & 
\colhead{$\log{\tau}$} 
}
\startdata
CoRoT-23 & $9.0_{-1.0}^{+1.0}$ & $9.2_{-1.5}^{+1.5}$ & $5900.0_{-100.0}^{+100.0}$ & $0.05_{-0.1}^{+0.1}$ & $4.01_{-0.08}^{+0.08}$ & $3.69_{-0.63}^{+1.39}$ & $1.9_{-0.14}^{+0.16}$ & $1.35_{-0.1}^{+0.08}$ & $7.56$ & $0.15$ \\ 
HATS-23 & $4.62_{-1.0}^{+1.0}$ & $...$ & $5780.0_{-120.0}^{+120.0}$ & $0.28_{-0.07}^{+0.07}$ & $4.33_{-0.04}^{+0.04}$ & $4.5_{-2.33}^{+2.18}$ & $1.09_{-0.05}^{+0.06}$ & $1.07_{-0.05}^{+0.04}$ & $8.21$ & $0.22$ \\ 
HD132406 & $2.16_{-1.0}^{+1.0}$ & $...$ & $5766.0_{-23.0}^{+23.0}$ & $0.14_{-0.02}^{+0.02}$ & $4.19_{-0.03}^{+0.03}$ & $7.6_{-1.06}^{+1.16}$ & $1.35_{-0.01}^{+0.01}$ & $1.07_{-0.04}^{+0.03}$ & $11.92$ & $8.47$ \\ 
HD149143 & $4.97_{-1.0}^{+1.0}$ & $28.0$ & $5950.0_{-21.0}^{+21.0}$ & $0.32_{-0.02}^{+0.02}$ & $4.21_{-0.04}^{+0.04}$ & $3.19_{-0.42}^{+0.44}$ & $1.69_{-0.02}^{+0.02}$ & $1.35_{-0.02}^{+0.03}$ & $8.23$ & $0.67$ \\ 
HD210277 & $2.29_{-1.0}^{+1.0}$ & $39.0$ & $5505.0_{-27.0}^{+27.0}$ & $0.18_{-0.02}^{+0.02}$ & $4.3_{-0.04}^{+0.04}$ & $6.74_{-1.7}^{+1.79}$ & $1.06_{-0.01}^{+0.01}$ & $1.01_{-0.03}^{+0.03}$ & $12.63$ & $8.68$ \\ 
HD23127 & $4.2_{-1.0}^{+1.0}$ & $33.0$ & $5891.0_{-33.0}^{+33.0}$ & $0.41_{-0.03}^{+0.03}$ & $4.23_{-0.05}^{+0.05}$ & $3.1_{-0.42}^{+0.45}$ & $1.61_{-0.02}^{+0.02}$ & $1.34_{-0.02}^{+0.02}$ & $13.12$ & $10.46$ \\ 
HD68988 & $3.62_{-1.0}^{+1.0}$ & $...$ & $5946.0_{-64.0}^{+64.0}$ & $0.34_{-0.05}^{+0.05}$ & $4.39_{-0.12}^{+0.12}$ & $1.37_{-0.81}^{+1.23}$ & $1.21_{-0.01}^{+0.01}$ & $1.23_{-0.04}^{+0.03}$ & $8.61$ & $1.8$ \\ 
HD9174 & $2.67_{-1.0}^{+1.0}$ & $...$ & $5631.0_{-30.0}^{+30.0}$ & $0.36_{-0.02}^{+0.02}$ & $4.05_{-0.04}^{+0.04}$ & $7.16_{-1.23}^{+0.54}$ & $1.68_{-0.02}^{+0.02}$ & $1.16_{-0.02}^{+0.07}$ & $13.3$ & $10.04$ \\ 
Kepler-1047 & $4.6_{-1.0}^{+1.0}$ & $31.85_{-2.45}^{+2.45}$ & $5658.0_{-60.0}^{+60.0}$ & $0.29_{-0.04}^{+0.04}$ & $4.23_{-0.1}^{+0.1}$ & $6.16_{-1.4}^{+1.92}$ & $1.59_{-0.09}^{+0.11}$ & $1.19_{-0.08}^{+0.08}$ & $14.87$ & $9.4$ \\ 
Kepler-1054 & $4.0_{-1.0}^{+1.0}$ & $19.33_{-1.65}^{+1.65}$ & $5909.0_{-60.0}^{+60.0}$ & $0.35_{-0.04}^{+0.04}$ & $4.18_{-0.1}^{+0.1}$ & $4.02_{-0.7}^{+1.13}$ & $1.51_{-0.11}^{+0.13}$ & $1.26_{-0.07}^{+0.06}$ & $12.41$ & $4.86$ \\ 
Kepler-1068 & $2.0_{-1.0}^{+1.0}$ & $17.58_{-0.1}^{+0.1}$ & $5684.0_{-60.0}^{+60.0}$ & $0.18_{-0.04}^{+0.04}$ & $4.5_{-0.1}^{+0.1}$ & $4.02_{-2.61}^{+3.17}$ & $1.02_{-0.06}^{+0.08}$ & $1.02_{-0.05}^{+0.05}$ & $13.39$ & $6.9$ \\ 
Kepler-111 & $2.8_{-1.0}^{+1.0}$ & $16.91_{-0.27}^{+0.27}$ & $5905.0_{-60.0}^{+60.0}$ & $0.22_{-0.04}^{+0.04}$ & $4.28_{-0.1}^{+0.1}$ & $4.39_{-1.67}^{+1.53}$ & $1.19_{-0.06}^{+0.07}$ & $1.11_{-0.04}^{+0.04}$ & $12.66$ & $4.8$ \\ 
Kepler-1141 & $3.3_{-1.0}^{+1.0}$ & $22.1_{-0.17}^{+0.17}$ & $5836.0_{-60.0}^{+60.0}$ & $0.1_{-0.04}^{+0.04}$ & $4.27_{-0.1}^{+0.1}$ & $6.25_{-1.69}^{+1.74}$ & $1.17_{-0.04}^{+0.05}$ & $1.04_{-0.04}^{+0.04}$ & $12.9$ & $4.66$ \\ 
Kepler-1211 & $2.9_{-1.0}^{+1.0}$ & $8.41_{-0.17}^{+0.17}$ & $5787.0_{-60.0}^{+60.0}$ & $0.1_{-0.04}^{+0.04}$ & $4.16_{-0.1}^{+0.1}$ & $7.74_{-1.78}^{+1.81}$ & $1.35_{-0.13}^{+0.16}$ & $1.06_{-0.06}^{+0.06}$ & $13.56$ & $6.45$ \\ 
Kepler-1269 & $4.5_{-1.0}^{+1.0}$ & $21.72_{-0.48}^{+0.48}$ & $5937.0_{-60.0}^{+60.0}$ & $0.15_{-0.04}^{+0.04}$ & $4.01_{-0.1}^{+0.1}$ & $4.42_{-1.03}^{+1.31}$ & $1.84_{-0.12}^{+0.15}$ & $1.28_{-0.09}^{+0.1}$ & $14.91$ & $9.16$ \\ 
Kepler-146 & $3.6_{-1.0}^{+1.0}$ & $13.85_{-0.48}^{+0.48}$ & $6000.0_{-60.0}^{+60.0}$ & $0.08_{-0.04}^{+0.04}$ & $4.41_{-0.1}^{+0.1}$ & $2.95_{-1.68}^{+2.06}$ & $1.11_{-0.05}^{+0.06}$ & $1.09_{-0.05}^{+0.05}$ & $13.8$ & $8.19$ \\ 
Kepler-1542 & $0.02_{-1.0}^{+1.0}$ & $43.52_{-5.12}^{+5.12}$ & $5544.0_{-60.0}^{+60.0}$ & $0.07_{-0.04}^{+0.04}$ & $4.26_{-0.1}^{+0.1}$ & $11.03_{-2.14}^{+1.64}$ & $1.18_{-0.05}^{+0.05}$ & $0.96_{-0.03}^{+0.04}$ & $13.38$ & $4.86$ \\ 
Kepler-157 & $1.07_{-1.0}^{+1.0}$ & $75.74_{-22.62}^{+22.62}$ & $5790.0_{-60.0}^{+60.0}$ & $-0.04_{-0.04}^{+0.04}$ & $4.34_{-0.1}^{+0.1}$ & $7.81_{-2.16}^{+2.31}$ & $1.12_{-0.07}^{+0.08}$ & $0.98_{-0.05}^{+0.05}$ & $12.19$ & $3.26$ \\ 
Kepler-170 & $1.08_{-1.0}^{+1.0}$ & $75.8_{-30.46}^{+30.46}$ & $5573.0_{-60.0}^{+60.0}$ & $0.39_{-0.04}^{+0.04}$ & $4.29_{-0.1}^{+0.1}$ & $6.82_{-2.79}^{+2.6}$ & $1.09_{-0.06}^{+0.07}$ & $1.03_{-0.04}^{+0.04}$ & $12.93$ & $5.48$ \\ 
Kepler-319 & $3.4_{-1.0}^{+1.0}$ & $13.42_{-0.03}^{+0.03}$ & $5598.0_{-60.0}^{+60.0}$ & $0.06_{-0.04}^{+0.04}$ & $4.65_{-0.1}^{+0.1}$ & $2.3_{-1.62}^{+2.54}$ & $0.89_{-0.03}^{+0.03}$ & $0.95_{-0.03}^{+0.03}$ & $12.88$ & $5.74$ \\ 
Kepler-36 & $4.9_{-1.0}^{+1.0}$ & $17.62_{-0.54}^{+0.54}$ & $5979.0_{-60.0}^{+60.0}$ & $-0.18_{-0.04}^{+0.04}$ & $4.11_{-0.1}^{+0.1}$ & $6.27_{-1.03}^{+0.9}$ & $1.64_{-0.08}^{+0.08}$ & $1.1_{-0.05}^{+0.06}$ & $13.47$ & $6.8$ \\ 
Kepler-422 & $2.8_{-1.3}^{+1.3}$ & $...$ & $5891.0_{-60.0}^{+60.0}$ & $0.21_{-0.04}^{+0.04}$ & $4.21_{-0.1}^{+0.1}$ & $4.64_{-0.91}^{+1.34}$ & $1.33_{-0.08}^{+0.08}$ & $1.16_{-0.05}^{+0.05}$ & $10.24$ & $2.95$ \\ 
Kepler-596 & $3.0_{-1.0}^{+1.0}$ & $68.58_{-2.99}^{+2.99}$ & $5983.0_{-60.0}^{+60.0}$ & $-0.06_{-0.04}^{+0.04}$ & $4.26_{-0.1}^{+0.1}$ & $5.99_{-1.39}^{+1.38}$ & $1.32_{-0.09}^{+0.11}$ & $1.07_{-0.05}^{+0.05}$ & $13.42$ & $7.02$ \\ 
Kepler-650 & $2.1_{-1.0}^{+1.0}$ & $26.96_{-0.55}^{+0.55}$ & $5848.0_{-60.0}^{+60.0}$ & $0.14_{-0.04}^{+0.04}$ & $4.15_{-0.1}^{+0.1}$ & $6.16_{-1.61}^{+1.49}$ & $1.57_{-0.18}^{+0.19}$ & $1.14_{-0.07}^{+0.1}$ & $11.91$ & $3.57$ \\ 
Kepler-750 & $4.4_{-1.0}^{+1.0}$ & $69.08_{-25.49}^{+25.49}$ & $5947.0_{-60.0}^{+60.0}$ & $-0.1_{-0.04}^{+0.04}$ & $4.15_{-0.1}^{+0.1}$ & $6.94_{-1.34}^{+1.35}$ & $1.41_{-0.11}^{+0.15}$ & $1.06_{-0.06}^{+0.06}$ & $12.93$ & $5.89$ \\ 
Kepler-772 & $2.6_{-1.0}^{+1.0}$ & $65.62_{-7.8}^{+7.8}$ & $5641.0_{-60.0}^{+60.0}$ & $0.12_{-0.04}^{+0.04}$ & $3.97_{-0.1}^{+0.1}$ & $7.03_{-1.17}^{+1.44}$ & $1.85_{-0.15}^{+0.18}$ & $1.14_{-0.07}^{+0.06}$ & $13.84$ & $6.73$ \\ 
Kepler-93 & $1.88_{-1.0}^{+1.0}$ & $57.32_{-1.3}^{+1.3}$ & $5624.0_{-40.0}^{+40.0}$ & $-0.15_{-0.03}^{+0.03}$ & $4.48_{-0.08}^{+0.08}$ & $6.42_{-3.0}^{+2.65}$ & $0.94_{-0.01}^{+0.01}$ & $0.92_{-0.04}^{+0.05}$ & $13.18$ & $9.79$ \\ 
WASP-133 & $1.56_{-1.0}^{+1.0}$ & $...$ & $5700.0_{-100.0}^{+100.0}$ & $0.29_{-0.12}^{+0.12}$ & $4.1_{-0.1}^{+0.1}$ & $6.12_{-1.33}^{+1.64}$ & $1.55_{-0.07}^{+0.07}$ & $1.18_{-0.07}^{+0.07}$ & $8.09$ & $-1.76$ \\ 
WASP-170 & $5.6_{-1.0}^{+1.0}$ & $7.75_{-0.02}^{+0.02}$ & $5593.0_{-150.0}^{+150.0}$ & $0.22_{-0.09}^{+0.09}$ & $4.0_{-0.2}^{+0.2}$ & $8.22_{-3.43}^{+3.05}$ & $1.05_{-0.03}^{+0.04}$ & $0.98_{-0.04}^{+0.06}$ & $8.4$ & $0.31$ \\ 
WASP-37 & $3.06_{-1.6}^{+1.6}$ & $21.0_{--9.0}^{+-9.0}$ & $5917.0_{-72.0}^{+72.0}$ & $-0.23_{-0.05}^{+0.05}$ & $4.45_{-0.15}^{+0.15}$ & $8.15_{-1.94}^{+2.1}$ & $1.08_{-0.04}^{+0.04}$ & $0.93_{-0.04}^{+0.04}$ & $8.32$ & $0.22$ \\ 
\enddata
\tablecomments{This is an excerpt of the table to illustrate its form and content. The entire table is available in the electronic form of the journal.  See Equations \ref{eq:eta} and \ref{eq:tau} for the definitions of $\eta$ and $\tau$.
}
\end{deluxetable*}

\subsection{Tidal Ranking}\label{subsec:tidal_rank}

To rank the systems according to the expected degree of tidal spin-up,
we used a metric based on the tidal theory described by \cite{Lai2012}.
In that theory, the timescale for tidal spin-up is
\begin{equation}\label{timescale}
    t_{\rm spin\mhyphen up} \approx  \frac{4Q'}{9}\,
    \frac{S}{L}\,
    \bigg(\frac{M}{m}\bigg)\bigg(\frac{a}{R}\bigg)^{\!5}
    \frac{1}{\Omega},
\end{equation}
where $Q'$ is the star's modified tidal quality
factor \citep{Goldreich1966}\footnote{According to this definition,
$Q' = 3Q/2k_2$, where $Q$ is the (un-modified) tidal quality factor,
and $k_2$ is the tidal Love number.},
$S$ and $L$ are the spin and orbital angular momenta,
$M$ and $m$ are the masses of the star and planet,
$a$ is the orbital radius (assuming a circular orbit),
$R$ is the stellar radius, and
$\Omega$ is the orbital angular frequency.
Using $S= \kappa MR^2\Omega_\star$ and
$L =  ma^2\Omega$,
we can rewrite this equation as
\begin{equation}\label{tspin}
t_{\rm spin\mhyphen up} \approx
\frac{4Q'\kappa}{9}
\left( \frac{M}{m} \right)^2 \left( \frac{a}{R} \right)^3
\left(\frac{\Omega_\star}{\Omega} \right) \frac{1}{\Omega}.
\end{equation}
We defined two dimensionless ratios
to rank the systems by the importance of
tidal effects. The first is a dimensionless
factor that appears in the preceding
equation,
\begin{equation}
\label{eq:eta}
  \eta \equiv \left( \frac{M}{m} \right)^2 \left( \frac{a}{R} \right)^3.
\end{equation}
The second dimensionless number is 
the ratio between the expected spin-up time
and the main-sequence age,
\begin{equation}
\label{eq:tau}
\tau \equiv \frac{t_{\rm spin\mhyphen up}}{{\rm age}}.
\end{equation}
For simplicity we assumed $\kappa = 0.06$ 
and $\Omega_\star = 2\pi R_\star/\vsini$ to allow $\tau$ to
be computed for all the stars in our sample.
With these definitions, lower values of $\tau$ or $\eta$
correspond to more opportunity for tidal spin-up.

\begin{figure}[ht!]
\centering
\includegraphics[width=0.45\textwidth]{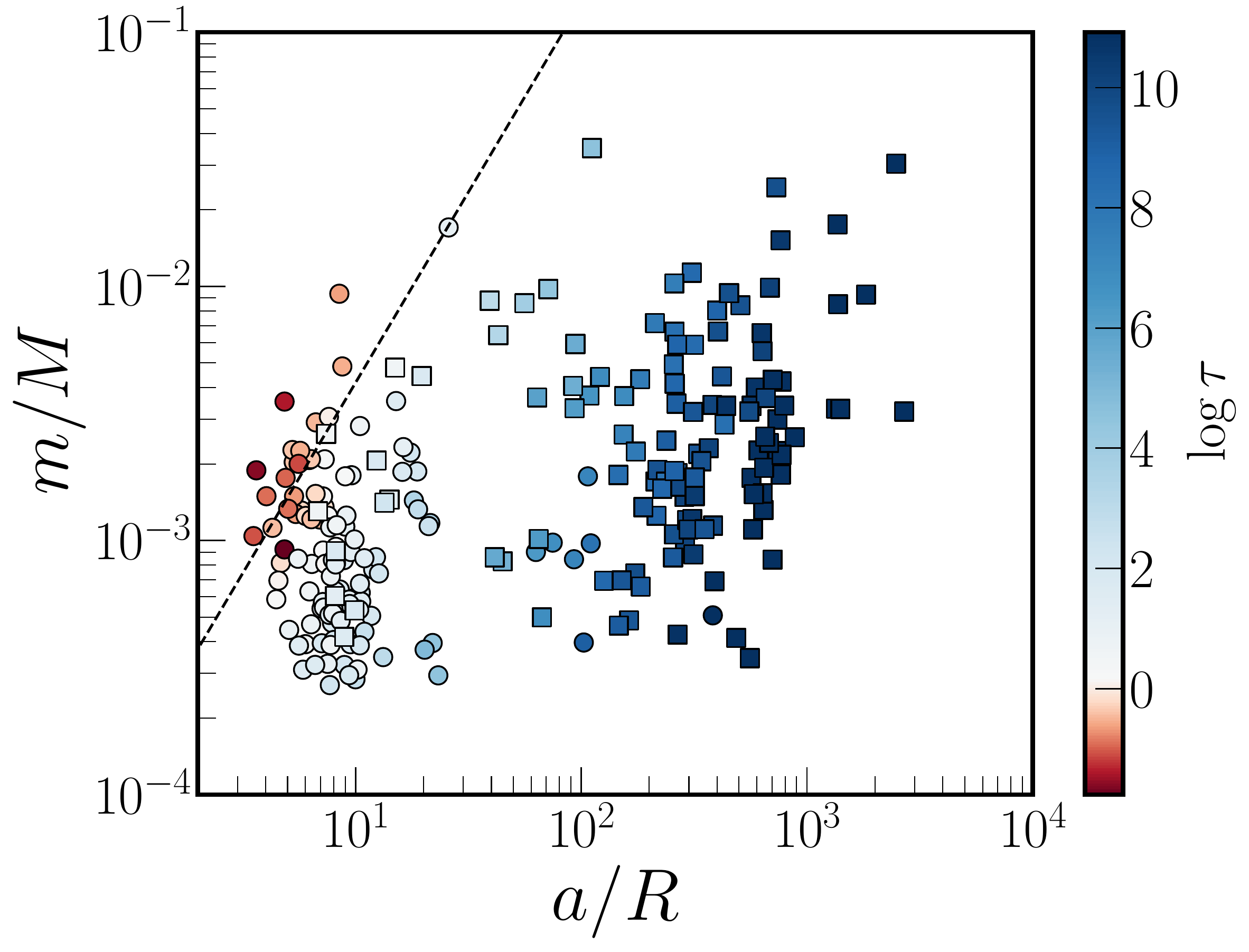}
\caption{Planet-to-star mass ratio ($m/M$)
and orbital distance to stellar radius ratio ($a/R$)
for stars with giant planets. Circles represent transiting planets,
and squares represent Doppler-detected planets.
The color conveys the value of $\log\tau$, where $\tau$
is the ratio of the
theoretical spin-up time to the estimated main-sequence age.
The blue (red) points depict systems for which we expect weak
(strong) tidal spin-up.
The white points are near the boundary of $\tau = 1.5$
chosen to separate
Hot Jupiters (HJ) and Control Jupiters (CJ).
This boundary is nearly equivalent to a critical value of
$1.6 \times 10^9$ for $\eta$, which is shown with the dotted line.
}
\label{fig:logeps_div}
\end{figure}

Figure \ref{fig:logeps_div} displays the distribution of $m/M$ and
$a/R$ in our sample, with color indicating $\log\tau$, and symbol
shape indicating whether the planet was detected with the Doppler
method (squares) or the transit method (circles).
Most of the low-$\tau$, low-$\eta$ systems are transiting planets, and most of the high-$\tau$, high-$\eta$ systems are Doppler planets --- as expected, given that transit surveys are more strongly
biased than Doppler surveys in favor of
close-orbiting giant planets.

Much of the subsequent analysis was based on the separation
of the samples into two groups, one of which is theoretically
expected to have experienced significant tidal
spin-up, and the other of which involves planets that are not massive
enough or close enough to the star to expect much tidal spin-up.
To divide the sample, we chose a critical value $\tau_{\rm c} = 1.5$, because it corresponds to a nominal case
in which the theoretical spin-up timescale is 10\,Gyr
for a Jupiter-mass planet in a 5-day orbit around a
Sun-like star with a rotation period of 25 days and
$Q' = 10^7$ at a corresponding age of 4.5\,Gyr.
For reference, this nominal case
also has $\eta = 1.6 \times 10^{9}$.

We refer to systems with $\tau<\tau_{\rm c}$ as Hot Jupiters (HJ) and
the giant-planet systems with $\tau>\tau_{\rm c}$
as Control Jupiters (CJ).
This definition led to a sample of 32 HJs and 208 CJs.
Among the CJs, 90 are
transiting planets and 118 were detected with the Doppler method.\footnote{Although the results described this paper were
obtained with the choice $\tau_{\rm c} = 1.5$,
we confirmed that none of our conclusions
hinge on this exact choice. Qualitatively similar results
are obtained for any value of $\tau_{\rm c}$ of
the same order of magnitude. We also obtained
similar results when dividing the samples
according to a critical value of $\eta$ instead
of $\tau$.}
These population numbers are also given in Table~\ref{tbl:hj_wj_populations}.
There was no need to divide up the CKS systems
because they all have $\tau\gg\tau_{\rm c}$,
given the low masses of the planets.

\begin{deluxetable}{cccc}
\tabletypesize{\footnotesize}
\tablecaption{Planet Samples
\label{tbl:hj_wj_populations}}
\tablecolumns{4}
\tablehead{
\colhead{Characteristic} &
\colhead{\hspace{10pt} HJ}\hspace{10pt} &
\colhead{CJ} &
\colhead{CKS}
}
\startdata
 Total number & 32 & 208 & 283 \\
 Transit Discoveries & 32 & 90 & 283 \\ 
 Doppler Discoveries  & 0 & 118 & 0 \\ 
 Photometric Rotation Periods & 19 & 24 & 67\\
 Spectroscopic Rotation Periods & 0 & 35 & 0
\enddata
\tablecomments{Characteristics of the hot and Control Jupiter samples.
Notably, all of the Hot Jupiters and CKS planets were detected in transit
surveys, while the Control Jupiters
contain a mixture of transiting and Doppler-detected planets.}
\end{deluxetable}

\begin{figure}[ht!]
\centering
\includegraphics[width=0.50\textwidth]{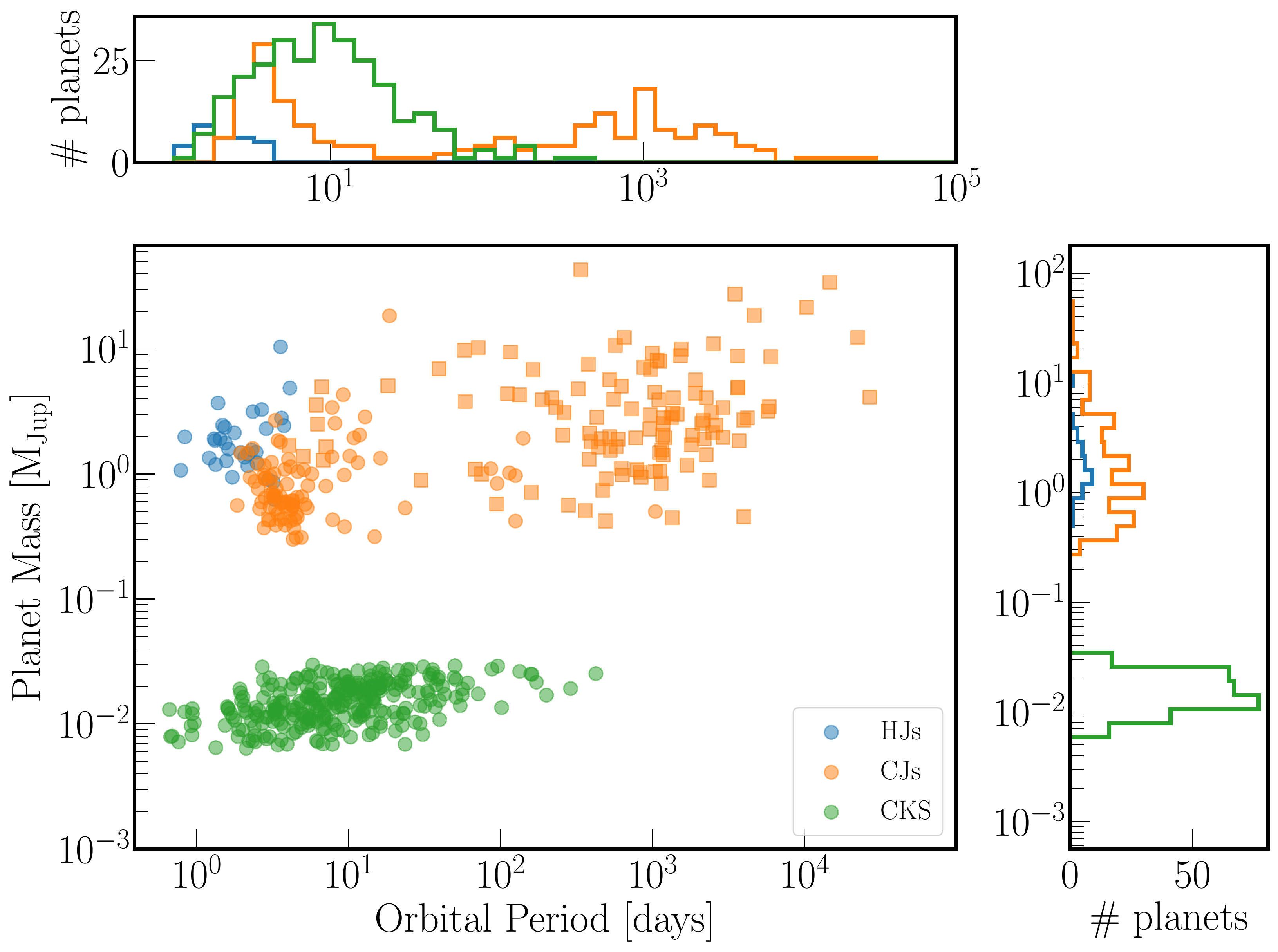}
\caption{Planet mass versus orbital period for the HJs, CJs, and CKS
planets. Circles represent transit-discovered planets; squares represent Doppler-discovered planets.}
\label{fig:mass_orbper}
\end{figure}

Figure~\ref{fig:mass_orbper} shows the distribution of masses and orbital periods of the planets in our samples.
The CKS period distribution overlaps with the HJ and CJ samples,
but the masses are all substantially lower.\footnote{We
calculated the expected planet masses based on their measured
radii, using Equation 1 of \cite{Wolfgang2016}.}
Thus, we have two control samples to compare to the HJ sample:
the CJs have similar masses and wider orbits,
while the CKS planets have smaller masses
and a broader range of orbital distances.

\subsection{Comparison of Spectroscopic Parameters}\label{subsec:spec_params}

Ideally, the control samples would consist of stars with the same
distribution of masses, compositions, and ages as the Hot Jupiter
hosts.  We should not expect them to align perfectly, because the
stars are drawn from different surveys and because of astrophysical
correlations between stellar and planetary properties (such as the
well-known tendency for giant-planet hosts to have
higher-than-average metallicity). Nevertheless, we can check
for any major mismatches that would invalidate our comparisons.

A concern with the Control Jupiters is that more than half
of the sample is drawn from Doppler surveys, whereas all of the Hot Jupiters were identified in transit surveys.
Transit and Doppler surveys are subject
to different selection effects favoring the detection of planets
around different types of stars. We must therefore make sure that
despite these different selection effects, 
the Control Jupiter hosts have spectroscopic properties similar
to those of the Hot Jupiter hosts.
Furthermore, the orbital inclinations of the transiting planets
are all very close to 90$^\circ$, while those of the Doppler
planets have a much broader range of inclinations.
This has two relevant
consequences.
First, the masses of the Doppler planets are formally unknown;
only the minimum possible mass ($m\sin i$)
can be measured. To account
for this ambiguity, whenever the planet mass was needed
we divided $m\sin i$ by $\pi/4$, the average value of $\sin i$
for random orientations.\footnote{In fact, for
a sample of Doppler-detected planets, $\langle \sin i\rangle > \pi/4$
because the sample is deficient in low-inclination systems,
but we neglected this minor effect.}
Second, to the extent that
the inclination of the stellar spin axis is correlated
with the orbital inclination, the transiting planets would have
a different distribution of $\vsini$ than the Doppler planets
even if they have the same distribution of rotation velocities.
We return to address this complication in
Sections~3 and \ref{sec:discussion}.

The CKS sample of small-planet host stars consists entirely of
transiting planets, so it does not suffer from the problems
just described for the giant-planet sample.  Here, a concern
is that giant-planet hosts are known to
be more metal-rich, as a whole, compared to the hosts
of smaller planets \citep{Gonzalez1997, Fischer2005, Santos2005, Petigura2018}.  Any astrophysical correlation
between metallicity and rotation would be a confounding factor in
the comparison between the CKS stars and the giant-planet hosts.

\begin{figure*}
\centering
\includegraphics[width=1.0\textwidth]{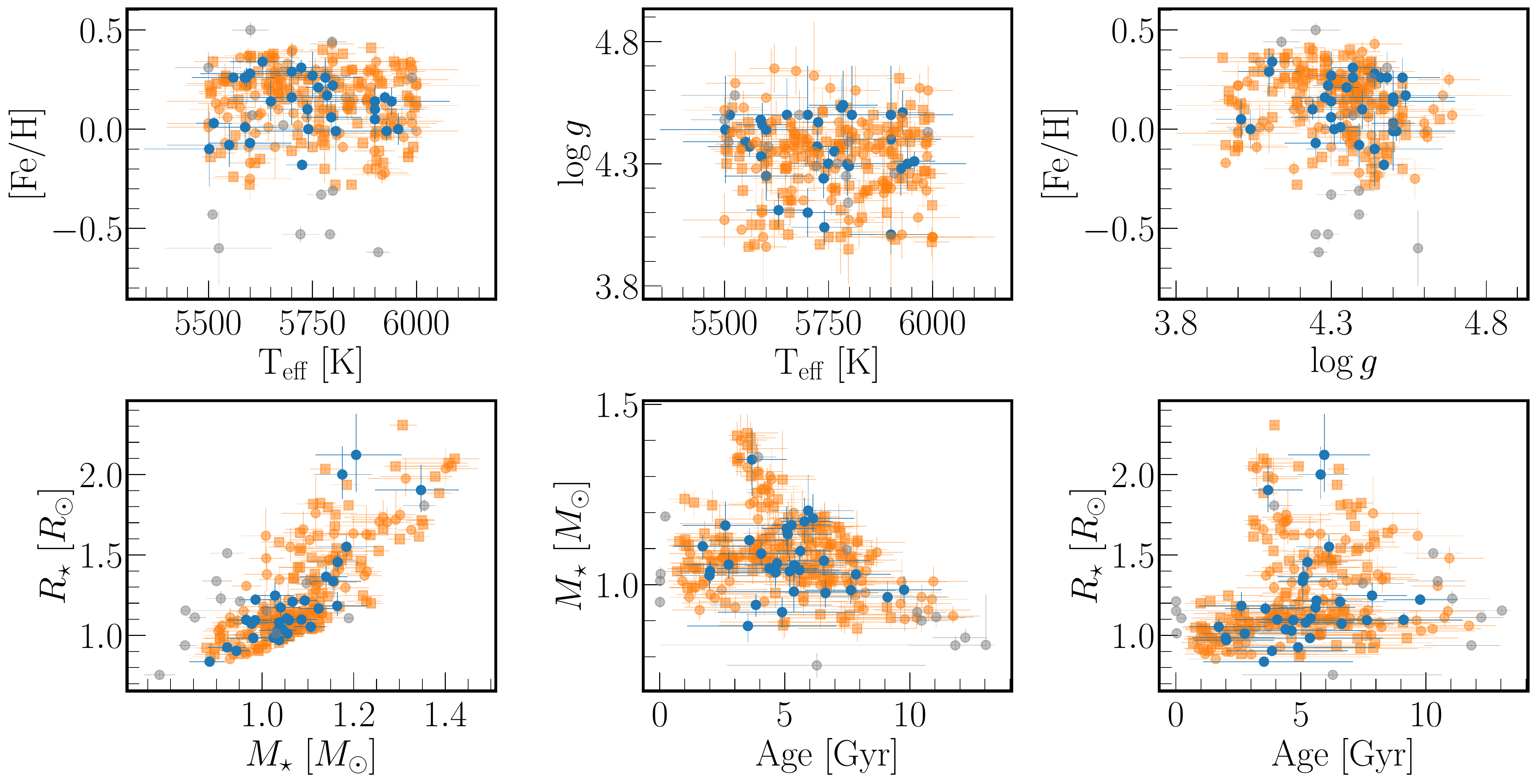}
\caption{Comparison of the spectroscopic and physical properties
of the host stars of Hot Jupiters (blue) and Control Jupiters (orange).
Circles are for transit detections and squares are for Doppler
detections. Note the absence of Hot Jupiters at low metallicities.
The gray points are those that were rejected for being outside
the designated metallicity range.
}
\label{fig:stellar_spec_params_wj}
\end{figure*}

\begin{figure*}
\centering
\includegraphics[width=1.0\textwidth]{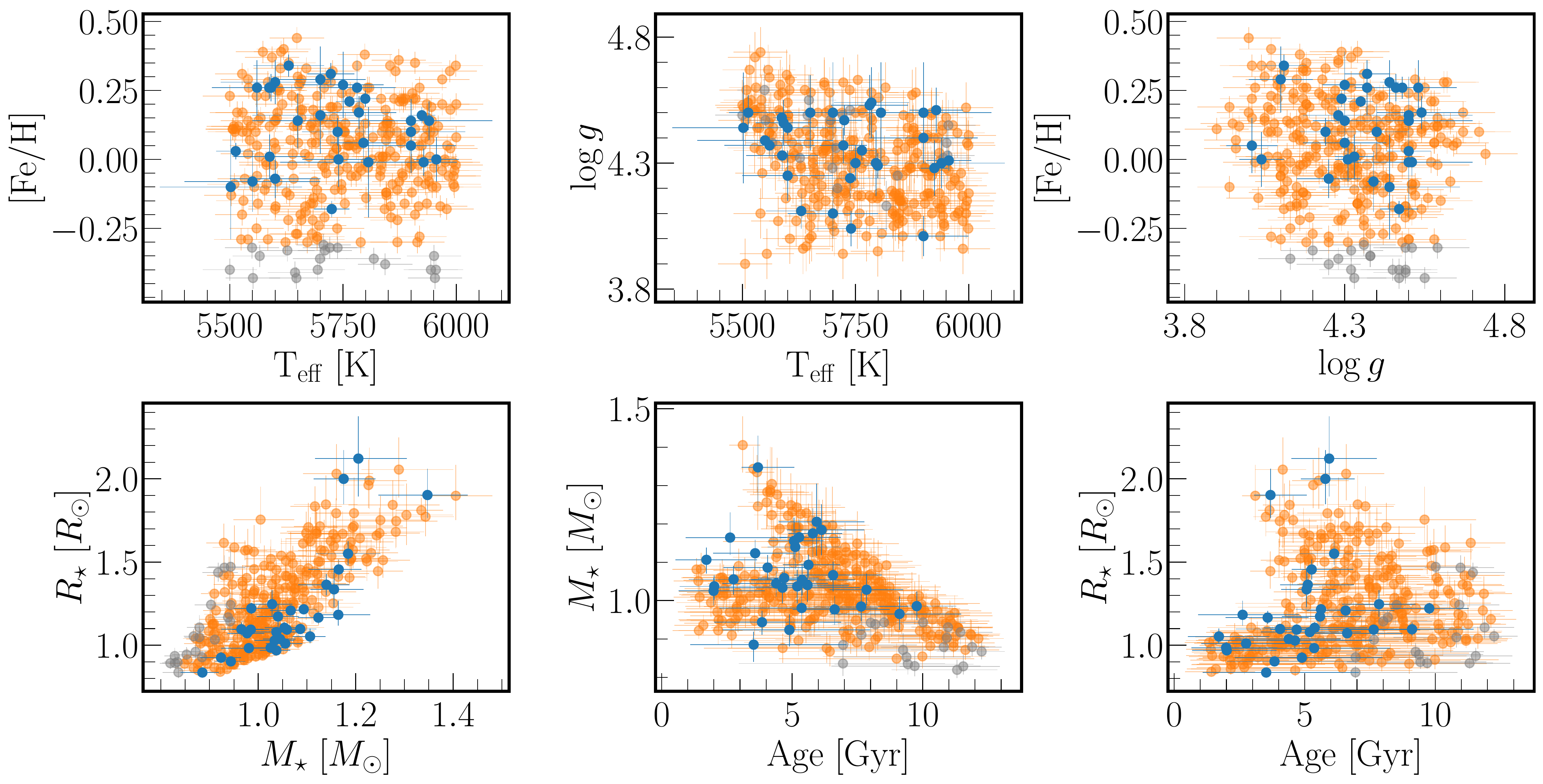}
\caption{Comparison of the spectroscopic and physical properties
of the host stars of Hot Jupiters (blue) and CKS planets (orange).
Circles are for transit detections and squares are for Doppler
detections.  The gray points are those that were rejected for being outside
the designated metallicity range.
In addition to the absence of Hot Jupiters at low metallicities,
note that the Hot Jupiters tend to be found around smaller,
higher-gravity, younger stars.
}
\label{fig:stellar_spec_params_cks}
\end{figure*}

Figures~\ref{fig:stellar_spec_params_wj} and  \ref{fig:stellar_spec_params_cks} compare the spectroscopic parameters and isochrone-fitting
results of the HJ, CJ, and CKS samples.
Our lower and upper limits on $\log g$ and [Fe/H] were determined
through inspection of these plots; the gray points are the stars that were rejected
because their parameters are too far afield from those of the majority
of the stars.
The metallicity effect 
is apparent in the upper right panel of Figure~\ref{fig:stellar_spec_params_cks}. 
There is also a tendency for the CKS stars to be assigned
smaller radii and older ages than the Hot Jupiter hosts of a given mass,
evident in the lower left panel of Figure~\ref{fig:stellar_spec_params_cks}. 
This could be related to the finding by \cite{Hamer2019} that the hosts
of hot Jupiters are kinematically younger (i.e.\ have a smaller
velocity dispersion) than similar stars without hot Jupiters. 

Apart from those patterns, the samples seem to span approximately
the same range of parameters.
We employed a two-sided Kolmogorov-Smirnov (KS) test for each spectroscopic parameter, to try and rule out the ``null hypothesis'' that the parameter values for the
HJ hosts and control stars are drawn from the same
distribution. The $p$-values, given in Table~\ref{tbl:pvals},
all exceed 0.05.

\begin{deluxetable}{cccc}
\tabletypesize{\footnotesize}
\tablecaption{$p$-values of two-sample
KS tests for the distributions of spectroscopic parameters relative to the HJ sample.
\label{tbl:pvals}}
\tablecolumns{4}
\tablehead{
\colhead{Control Sample} &
\colhead{\hspace{10pt} $\teff$}\hspace{10pt} &
\colhead{Metallicity} &
\colhead{$\log g$}
}
\startdata
 CJ  & 0.36 & 0.79 & 0.14 \\ 
 CKS & 0.43 & 0.07 & 0.08
\enddata
\end{deluxetable}

\pagebreak
\section{Comparison of Rotation Parameters}\label{sec:results}

We gathered all of the available information about the rotation
properties of the stars.
We found $\vsini$ measurements for all the stars in the literature.
We decided to regard as upper limits all the cases in which $v\sin i$ was reported
to be lower than 2\,km\,s$^{-1}$, out of concern about systematic errors.

We also searched the literature for stellar rotation periods measured
from either time-series broadband photometry, or spectroscopic monitoring
of emission-line fluxes.
We did not accept rotation periods based on measurements
of $v\sin i$ and the assumption $\sin i = 1$,
nor did we use estimated rotation periods based only on
the overall level of chromospheric activity. Needless to say,
we did not use rotation periods based on gyrochronology, either.
Using these criteria, we found
rotation periods for 78 stars (19 Hot Jupiters and 59 Control Jupiters). 
We also found measurements of the photometric
rotation periods for 68 of the CKS stars
in the catalog of
\cite{Mazeh2015}.\footnote{We only considered the
photometric rotation periods that were considered
most reliable by \cite{McQuillan2014}; in their terminology,
the ``weight'' exceeds 0.25.
We also decided to omit Kepler-1563 because the reported rotation
period of 46 days is nearly twice as long as any of the other rotation
periods in the sample. We believe that measurements of such
long periods are subject to
extra uncertainty because the Kepler data segments (``quarters'') span only 90 days.}
We adopted the period uncertainties from the literature,
and assumed an uncertainty of $20\%$ when the uncertainty was
not clearly documented. 

\subsection{Projected rotation velocity}
\label{subsec:vsini}

We compared the $v\sin i$ distributions of our samples to see if the HJ hosts
are rotating systematically faster, given their ages.  However, any observed differences
in the $\vsini$ distributions could be due to differences in $\sini$
as well as rotation velocity.  It seems safe to assume that the hosts
of all the Doppler-detected planets are nearly randomly-oriented, in which case any systematic differences in $\vsini$ between large populations of stars can be interpreted as differences in rotation velocity.

The situation is more complicated for the hosts of transit-detected planets. Transiting planets all have orbital
inclinations near $90^\circ$. Thus, if the stellar rotation
axes are aligned with the orbital axes, the stars with
transiting planets would have systematically higher $v\sin i$ values
than stars with Doppler planets, even if their distributions
of rotation velocities were the same. This complicates our
comparison between HJs and CJs, because the HJs are composed
mainly of transit-detected planets while the CJs are a more even
mixture of transit and Doppler discoveries.

Previous measurements have established that Sun-like
stars with hot Jupiters generally have low obliquities
implying $\sin i \approx 1$ \citep[see, e.g.,][]{Winn2010,Albrecht2012}.
The obliquity distribution of stars
with smaller or wider-orbiting planets is less well known.
There is statistical evidence that the Sun-like stars in the
{\it Kepler} sample (including the stars in our CKS sample)
have low obliquities,
based on the distribution of photometric variability
amplitudes \citep{Mazeh2015} as well as tests involving
$\vsini$ data \citep{Winn2017}.
These results suggest that
the $\sini$ distributions of the transiting Hot Jupiters and the CKS stars
are similar, and it is safe to interpret any systematic
differences in $v\sin i$ as differences in rotation velocity.

However, for the Control Jupiters, things may be different.
There are known cases of Sun-like stars with high obliquities
relative to wider-orbiting giant planets, such as 
WASP-17b, WASP-130b, WASP-134b, and HATS-18b \citep[][respectively]{Smith2013, Hellier2017, Anderson2018, Brahm2016}.
Hence, the mean value of $\sin i$ in the sample of transiting HJs
may be higher than that of the transiting CJs, confounding
the interpretation of the differences in the $v\sin i$ distributions.

We dealt with this issue by considering two limiting
cases. In the first case, the transiting HJs and CJs were both assumed to have
low obliquities and $\sin i =1$.  In the second case, which we consider
rather extreme, 
the transiting HJ hosts were assumed to have zero obliquity ($\sin i =1$)
and the transiting CJ hosts were also assumed to be randomly oriented
($\sin i=\pi/4$ on average).

\begin{figure*}[ht!]
\centering
\includegraphics[width=1.0\textwidth]{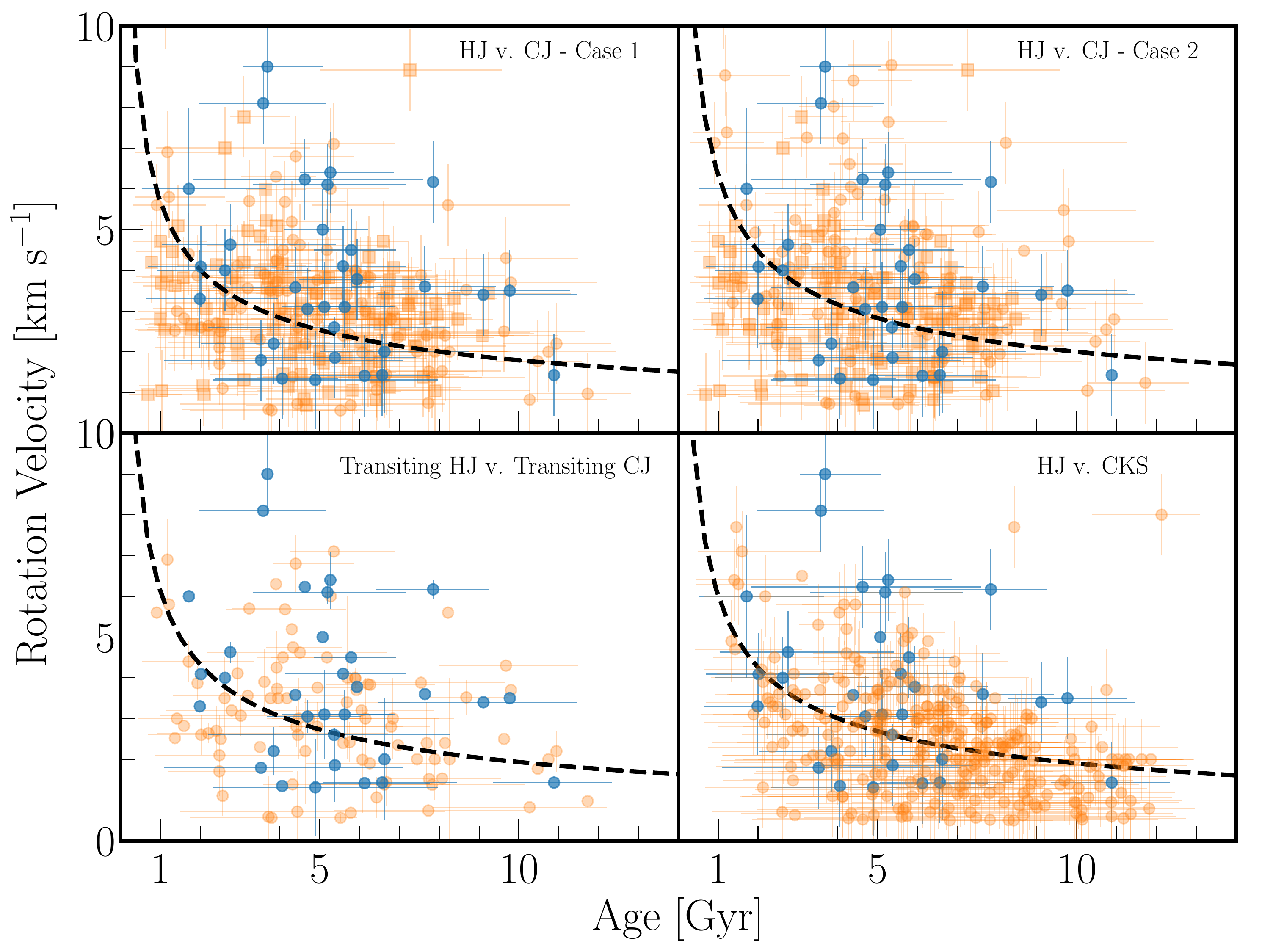}
\caption{Rotation velocity as a function of isochrone age. Blue points are for HJ hosts
and orange points are for control stars. Squares depict Doppler-discovered
planets, and circles depict transit-discovered planets. The dashed line is the best fit
to the control stars using Equation~\ref{sku_vsini}.
The plotted ``rotation velocity'' is $\vsini$ for the transiting HJs, $\vsini$
for the CKS stars, and $\vsini/(\pi/4)$ for the Doppler-detected planets. For the
transiting CJs, two different cases are considered.
{\it Top left.}---Case 1, assuming transiting CJs have $\sini = 1$.
{\it Top right.}---HJs versus CJs, assuming the CJ hosts are randomly oriented
($\langle\sin i\rangle=\pi/4$).
{\it Bottom left.}---Transiting Hot and Control Jupiters only assuming CJ hosts are randomly oriented. 
{\it Bottom right.}---HJs versus CKS stars.
The two outliers are Kepler-1505 and Kepler-461, both of which also have unusually
high variability amplitudes.}
\label{fig:vsini_results_comp}
\end{figure*}

Figure~\ref{fig:vsini_results_comp} compares the rotation-velocity distributions
of the HJs and control stars. In all of the panels, the plotted velocity
for the Doppler planets is $\vsini$ divided by $\pi/4$.
The top left panel represents the first case described above:
for all the transiting planets, the plotted velocity is $\vsini$.
The top right panel represents the second case:
the plotted velocity is $\vsini$ for the transiting HJs, and $\vsini/(\pi/4)$
for all CJs. 
The bottom right panel compares the HJ and CKS samples;
in this case the plotted velocity is $\vsini$ for all the transiting
planets.
Finally, the bottom left panel of Figure~\ref{fig:vsini_results_comp} compares the HJs and CJs after
excluding the Doppler planets; thus, it is a transit-to-transit comparison.
For this figure, the plotted velocity is $\vsini$ for
both HJs and CJs.

In all of these cases, the HJ hosts appear to have systematically faster
rotation than the CJ hosts.
To quantify the differences, we fitted a \citet{Skumanich1972} law,
\begin{equation}\label{sku_vsini}
     v\sin{i} = v_0~\bigg(\frac{\mathrm{age}}{5\,\mathrm{Gyr}}\bigg)^{-1/2},
\end{equation}
to the control-star data.  The only free parameter was $v_0$.
The HJ data fall mainly above the best-fitting curves
(i.e., the blue points are mainly above the dashed lines).
We defined a sum-of-residuals statistic,
\begin{equation}\label{residual}
    S = \sum_{n=1}^N \left(v\sin i_{n,{\rm obs}} - v\sin i_{n,{\rm calc}}\right),
\end{equation}
where $v\sin i_{\rm obs}$ is the observed value,
$v\sin i_{\rm calc}$ is the calculated value using the best-fitting
function (Equation~\ref{sku_vsini}), and the sum runs over all the data points. To estimate the probability that high $S$ values are the result
of random fluctuations, we used a Monte Carlo procedure.
In each Monte Carlo realization of the data,
we randomly drew (with replacement) a subset of stars from the entire
sample of stars --- both Hot Jupiters and control stars ---
to play the role of fictitious Hot Jupiters.
We performed $10^6$ such simulations and asked how often
the $S$ value of the simulated data was at least as large as the $S$ value
of the real data. The resulting $p$-values, given in the first column
of Table~\ref{tbl:res_pvals}, are less than 0.004
regardless of the control sample or assumed obliquity
distribution. These low $p$-values confirm the visual impression that the HJ hosts
tend to rotate faster at all ages.

\begin{deluxetable}{ccccc}
\tabletypesize{\footnotesize}
\tablecaption{$p$-values of two-sample
KS tests for the distributions of rotation parameters relative to the HJ sample.
\label{tbl:res_pvals}}
\tablecolumns{5}
\tablehead{
\colhead{Control Sample} &
\colhead{$\vsini$, case 1} &
\colhead{$\vsini$, case 2} &
\colhead{$P_{\rm rot}$}
}
\startdata
 CJ  & $2\times 10^{-4}$   & $4\times 10^{-3}$ & $\lesssim 10^{-6}$  \\ 
 CKS & $3\times 10^{-4}$ & $3\times 10^{-4}$ & $8\times 10^{-5}$ 
\enddata
\tablecomments{Case 1 assumes the transiting CJs have $\sin i=1$, while
Case 2 assumes they have $\langle \sin i\rangle= \pi/4$.}
\end{deluxetable}

\begin{figure*}[ht!]
\centering
\includegraphics[width=1.0\textwidth]{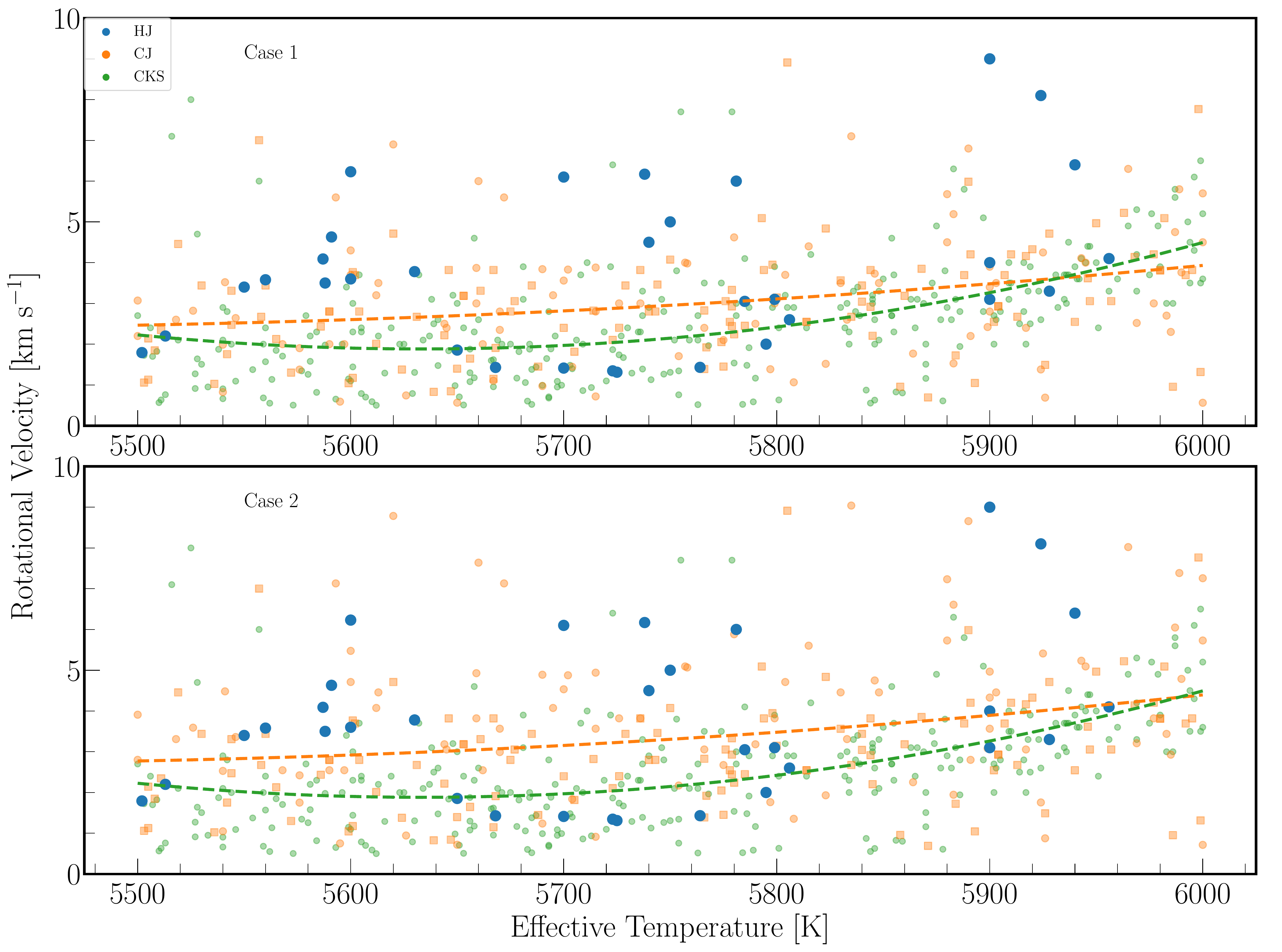}
\caption{Rotation velocity as a function of effective temperature,
for HJ hosts (blue points), CJs (orange points), and CKS stars (green points).
As in Figure~\ref{fig:vsini_results_comp}, the plotted velocity is
$\vsini$ for the transiting HJs and CKS stars, and $\vsini/(\pi/4)$ for the Doppler
planets. Two cases are considered for the transiting CJs.
{\it Top.}---Case 1, in which the plotted velocity is $\vsini = 1$ for CJ transit discoveries.
{\it Bottom.}---Case 2, in which it is $\vsini/(\pi/4)$.
The dashed curves are quadratic functions fitted to the control-star data.
\label{fig:vsini_teff}}
\end{figure*}

Another way to see the evidence for spin-up is to examine
the $\vsini$ distribution as a function of effective temperature
rather than isochrone age, as shown in Figure~\ref{fig:vsini_teff}.
This comparison has the advantage of being
independent of uncertainties in the isochrone
ages.
At any given effective temperature, the Hot Jupiter hosts
have systematically higher $\vsini$ values than the stars
in the control samples.
Following the example of \cite{Louden2020},
we fitted a quadratic function to the relationship between $\vsini$ and $\teff$ for each control sample (dashed curves, in Figure~\ref{fig:vsini_teff}).
The residual test discussed above yields a $p$-value $\lesssim 10^{-6}$ for the comparison
with CKS stars.  For the comparison with the CJs, the $p$-value
is $5.0 \times 10^{-5}$ and
$3.4 \times 10^{-3}$ for Cases 1 and 2, respectively.

\subsection{Rotation period}
\label{subsec:prot}

By comparing rotation periods rather than projected rotation velocities,
we avoid the complications due to the unknown obliquity distributions.
The penalty, though, is that rotation periods have only been measured
for a subset the stars in our sample.  This reduces the
sample sizes and the statistical power of any comparisons.
Furthermore, the stars with measured rotation
periods are not necessarily representative of the whole sample.
Rotation periods are easier to measure when the amplitude of
variability is high, which in turn is associated
with youth, rapid rotation, and high inclination.
While these biases should apply to all of the samples and
cancel out to some degree, there may be residual biases
that are difficult to quantify.

\begin{figure*}[ht!]
\centering
\includegraphics[width=0.45\textwidth]{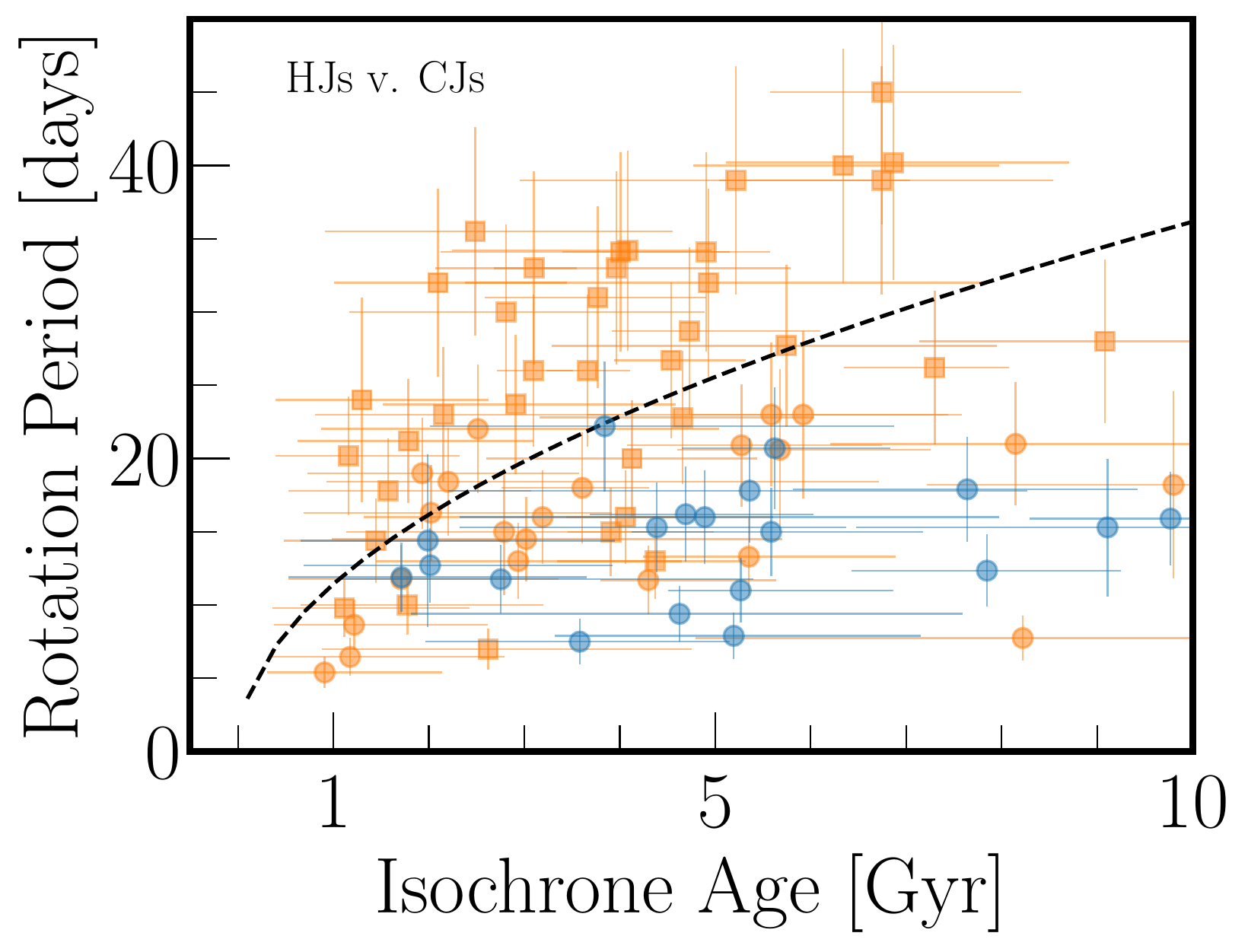}
\includegraphics[width=0.45\textwidth]{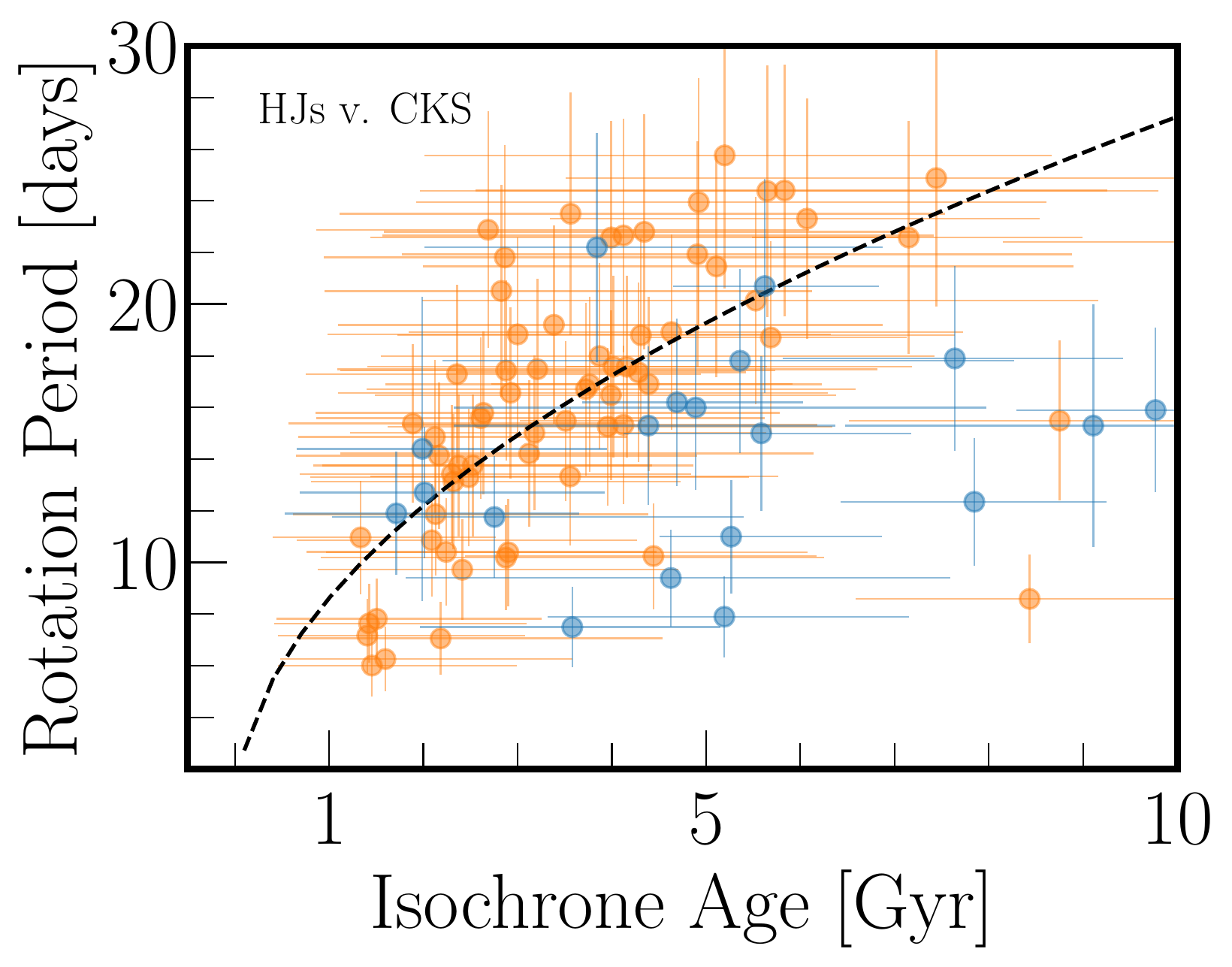}
\caption{Rotation period versus isochrone age for the
HJs (blue) and the control sample (orange), which is the CJ sample
in the left panel and the CKS sample in the right panel.
Squares are for Doppler-detected planets and circles
are for transit-detected planets.
The dashed curves are functions with the form of Equation~\ref{sku_prot}
fitted to the control sample.
}
\label{fig:prot_results}
\end{figure*}

These limitations notwithstanding, Figure~\ref{fig:prot_results}
shows rotation period versus isochrone age for the HJ, CJ, and CKS samples.
Compared to the control samples,
the Hot Jupiter hosts have systematically shorter periods
and more rapid rotation.  The left panel, in which the CJs are the
control sample, shows that Doppler-detected
planets tend to have longer periods than transit-detected planets.
This could be due at least in part to a bias against rapid rotators
in the Doppler surveys.  However, the right panel, in which the CKS stars are
the control sample, does not suffer from that particular bias
and also shows the HJ hosts to be rotating faster.

As before,
we quantified the differences by fitting the control data
to a Skumanich-like law,\footnote{We also experimented
with  more sophisticated and mass-dependent functions relating period and age
taken from \cite{Delorme2011a}, \cite{Cameron2009}, and \cite{Angus2019} and in all cases reached similar conclusions as those described
in this Section.}
\begin{equation}\label{sku_prot}
  P_{\rm rot} = P_0\bigg(\frac{\mathrm{age}}{5\,\mathrm{Gyr}}\bigg)^{1/2},
\end{equation}
\noindent where $P_0$ is a free parameter,
and then calculated the period-based
sum-of-residuals $S$ for the HJ host stars.
In this case, the HJ hosts have a negative value of $S$, i.e., they
have shorter rotation periods than would be predicted from the fit
to the control-star data.
The $p$-values, given in Table~\ref{tbl:res_pvals},
are low enough for the pattern to be deemed highly significant.

\subsection{Rotation vs.\ tidal spin-up parameter}\label{subsec:cont_tides}

The preceding tests convinced us that the HJ hosts do indeed rotate systematically
faster than the control stars, whether the measure of rotation is the projected
rotation velocity or the rotation period, or whether the comparison
is performed as a function of isochrone age or effective temperature.
To search for evidence that tidal spin-up is the reason for the excess
rotation, we tested for a correlation between rotation
and the theoretically expected degree of tidal spin-up, quantified
by our $\tau$ parameter. As a reminder, $\tau$ is the ratio of
the tidal spin-up time scale in the theory of \cite{Lai2012} and
the isochrone age, with lower values corresponding to a higher
expectation for tidal spin-up.

\begin{figure*}[ht!]
\centering
\includegraphics[width=1.0\textwidth]{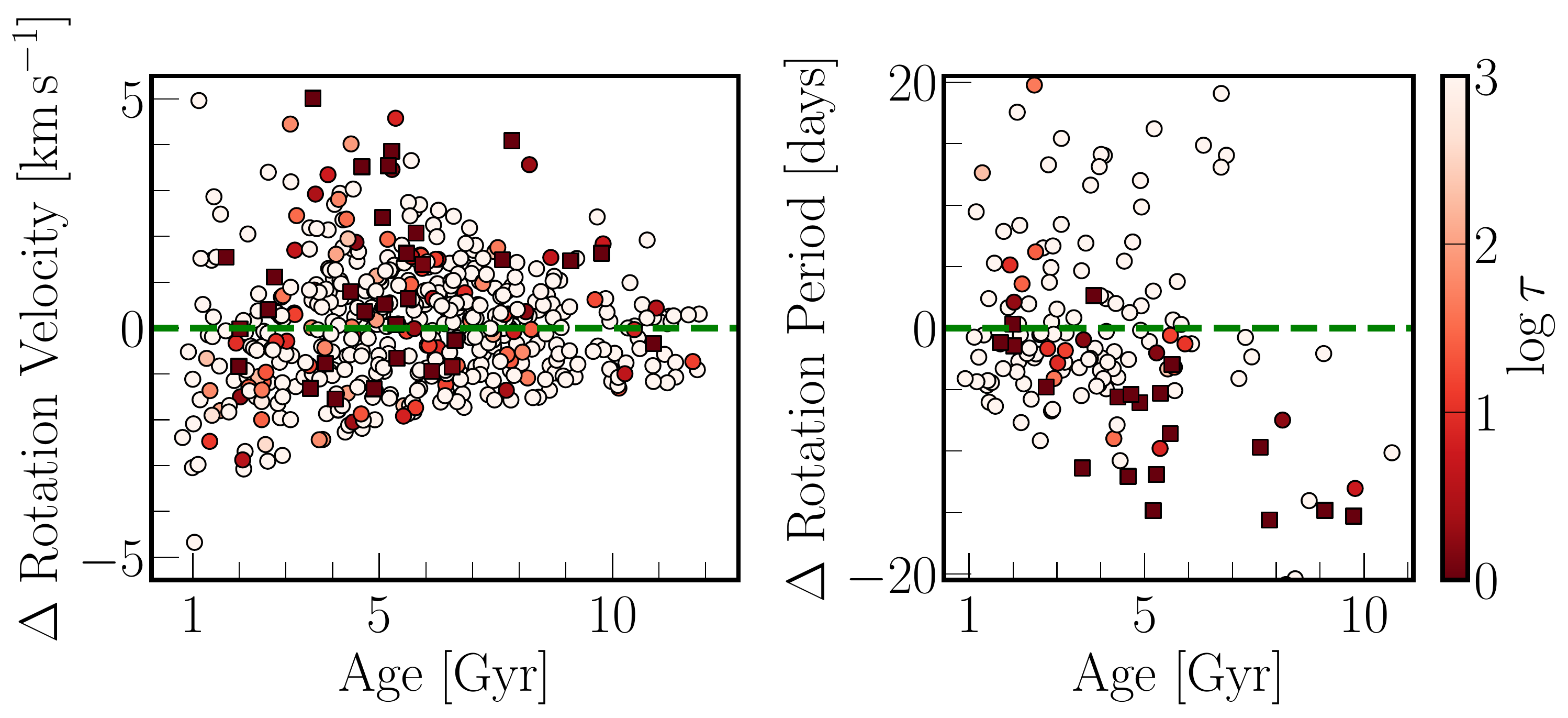}
\caption{Case 1 deviation of rotation velocity and rotation period as a function of isochrone age,
color-coded according to $\tau$, the tidal spin-up timescale divided by the isochrone age. Green fiducial line indicates $\Delta \vsini = 0$.
The data from all the samples are plotted (HJ, CJ, and CKS). For clarity in this particular figure, the squares represent the Hot Jupiter planets and circles illustrate the Control Jupiters. Smaller $\tau$ values, indicating strong tides, show higher excess velocity from the Skumanich Law.
\label{fig:vsini_tau_cbar}}
\end{figure*}

\begin{figure*}[ht!]
\centering
\includegraphics[width=1.00\textwidth]{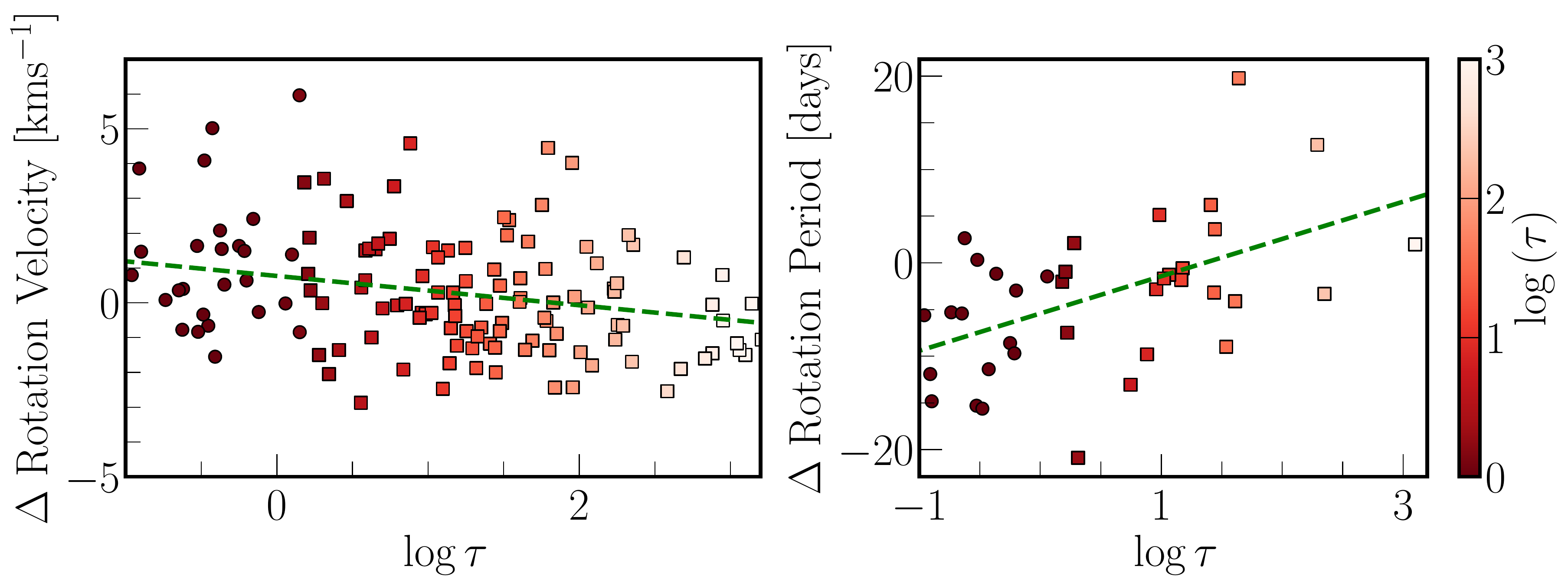}
\caption{Projected rotation velocity (left) and rotation period (right) as a
function of $\tau$, the tidal spin-up timescale divided by the isochrone age.
Stars with lower values of $\tau$ show faster rotation velocities and
shorter rotation periods. Squares represent Doppler-detected planets and circles depict transit-detected planets. 
\label{fig:tau_correlation}}
\end{figure*}

Figure~\ref{fig:vsini_tau_cbar} reproduces the data that were shown in
Figures~\ref{fig:vsini_results_comp} and \ref{fig:prot_results}, but
in this case the color of each point conveys the calculated
value of $\tau$, with darker points representing systems
where tidal spin-up should be most significant.
As expected, the darker points tend to be associated
with higher rotation
velocities and shorter rotation periods.
Figure~\ref{fig:tau_correlation} shows more directly the association
between excess rotation and $\tau$.
For this figure, the excess
rotation was defined as the difference between the observed value
of rotation velocity or period, and the calculated value based on
the Skumanich-like function fitted to the control-star data.
The plots are restricted to the range of $\tau$ between
0.1 and 1000, where we might expect to see a correlation
(this excludes many of the CKS systems for which $\tau\gg 1000$).

The excess rotation and the parameter $\tau$ are indeed correlated,
as confirmed through least-squares fitting.
Between $\vsini$ and $\tau$ there is a shallow but significant negative correlation:
Table~\ref{tbl:corr_tests} gives the results
of the Pearson and Spearman tests.\footnote{The Pearson
correlation coefficient is the covariance of two
variables divided
by the product of their standard deviations. The Spearman rank correlation coefficient is the Pearson correlation
coefficient between the rank-ordered values of the
two variables.} 

\begin{deluxetable}{ccccc}
\tabletypesize{\footnotesize}
\tablecaption{Correlation coefficients ($r$) and
$p$-values between $\tau$ and rotation parameters.\label{tbl:corr_tests}}
\tablecolumns{5}
\tablehead{
\colhead{Parameter} &
\multicolumn{2}{c}{Pearson} &
\multicolumn{2}{c}{Spearman} \\
\colhead{} &
\colhead{$r$} &
\colhead{$p$} &
\colhead{$r$} &
\colhead{$p$}
}
\startdata
 $\vsini$  & $-0.26$& $2.6\times 10^{-3}$ & $-0.24$ & $6.0 \times 10^{-3}$ \\ 
 $P_{\rm rot}$  & $+0.53$ &$4.6\times 10^{-4}$ &$+0.50$ & $1.0\times 10^{-3}$
\enddata
\end{deluxetable}

\section{Summary and Discussion}
\label{sec:discussion}

We investigated the evidence for tidal spin-up in hot Jupiter systems,
by comparing the rotation velocities and spin periods of Sun-like stars with a wide range of ages and planet parameters.
The stars that are theoretically expected to have been most susceptible to tidal spin-up ---
those with close-orbiting giant planets --- are indeed rotating faster than comparable stars
with other types of planets.  
By preparing appropriate samples and performing simple comparisons,
our approach was intended to be more empirical and less model-dependent than
the complementary studies of \cite{Jackson2009, Hansen2010, Ferraz-Mello2015a, Penev2018, Barker2020}
and \cite{Anderson2021}, who have modeled secular evolution in the context
of specific tidal models. In our study, the only input from theory was
in the definition of the dimensionless parameters
$\eta$ and $\tau$
that quantify the expected degree of tidal effects.

A limitation of our study is that 
although we did perform the isochrone analysis for all the stars
in the same manner,
the input data were all drawn from the literature, which means
the spectroscopic and rotation parameters were derived by different authors using different
techniques. The CKS parameters were homogeneously derived, but the giant-planet parameters
come from heterogeneous sources. Another limitation is that our samples contain planets discovered
in surveys with different selection biases. All of the HJs and
CKS planets were discovered with the transit technique, while the CJs consist of a mixture
of transit-detected and Doppler-detected planets. The Doppler surveys do not find
many planets around young and rapidly rotating stars because of the difficulty
of achieving good precision when the spectra have broad lines with time-variable
distortions.  There is also the issue that the distribution of $\sini$ is different
for the Doppler and transit-detected systems, and may also vary with the planet
properties. 
We dealt with these issues by performing additional tests with different subsamples
(e.g., only transiting planets) and under different assumptions about the
$\sini$ distribution (Cases 1 and 2). These robustness tests led to the same conclusion --- the HJ hosts
spin faster than the control stars --- but, naturally, with reduced
statistical significance.

Our results, based on larger samples and more controlled comparisons,
reinforce earlier evidence presented by \cite{Brown2014} and \cite{Maxted2015}
that close-orbiting giant planets are able to influence the
rotation rates of their host stars while they are on the main sequence.
Besides measurements of rotation rates,
another line of evidence that pointed to the same conclusion was presented
by \cite{Poppenhaeger2014}, who found several hot Jupiter hosts to
be more chromospherically active than their wide-orbiting binary companions.

An immediate implication is that gyrochronology is unreliable for stars with hot Jupiters, in agreement with previous empirical
findings by \cite{Brown2014} and \cite{Maxted2015}, and theoretical work by \cite{Ferraz-Mello2016}. 
The spin history of these stars is abnormal, invalidating the usual relationships
between mass, age, and rotation velocity.
Another implication is that hot Jupiter orbits decay significantly during
the main-sequence lifetime of the star; the angular momentum that is transferred
to the star's rotation must come
at the expense of the planet's orbit.
The same conclusion was reached by 
\cite{Hamer2019}, who showed that Sun-like stars with Hot Jupiters
are ``kinematically young,'' i.e., they
have a lower Galactic velocity dispersion than
similar stars without Hot Jupiters. They took the low
occurrence of Hot Jupiters around kinematically older stars
to be evidence for tidal destruction on Gyr timescales.
Our work supports this conclusion not only by identifying excess rotation of the HJ hosts,
but also in observing the tendency for HJ hosts to have
younger isochrone ages (Figures~\ref{fig:stellar_spec_params_wj} and  \ref{fig:stellar_spec_params_cks}).

The evidence for tidal transfer of angular momentum also suggests that 
Hot Jupiters can affect the spin direction of the star and that obliquity damping
can occur while the star is on the main sequence.  In this sense,
our results complement earlier work showing that Sun-like stars
with Hot Jupiters tend to have low obliquities, while stars that
are more massive or that have wider-orbiting planets are
sometimes observed to have high obliquities \citep{Winn2010,Albrecht2012}.
Theorists have indicated that
the timescales for spin-up (and the associated orbital decay) and obliquity
alteration need not always be the same \citep{Lai2012,Barker2020}, but for
Sun-like stars with Hot Jupiters, both effects do appear to be significant.

It would be interesting to extend this study to other types of stars,
both less and more massive than the Sun-like stars considered here,
because the mechanisms for tidal dissipation may be quite
different.  For massive stars, at this stage the main difficulty would be constructing
suitable control samples. There are plenty of F stars known to have
Hot Jupiters, but not as many F stars with wide-orbiting giant planets
or considerably smaller planets.  For the extension to low-mass stars, one problem is
that Hot Jupiters are themselves rare around low-mass stars. Moreover, our understanding of the rotational evolution of low-mass stars is limited in comparison to our understanding of Sun-like stars.  The situation will probably
improve as the NASA TESS mission \citep{Ricker2014} continues to find planets
and measure rotation periods
with ever greater sensitivity \cite[see, e.g.][]{Martins2020}, and after
the PLATO mission commences \citep{Rauer2014}.

\software{Astropy \citep{AstropyColab2018},
          Jupyter Notebooks \citep{Kluyver2016},
          Matplotlib \citep{Hunter2007},
          NumPy \citep{numpy2011, Harris2020},
          Pandas \citep{reback2020pandas},
          SciPy \citep{Virtanen2020},
          VizieR \citep{Ochsenbein2000},
}
 
\acknowledgements We thank Kaloyen Penev for helpful discussions.
This work was supported by NASA grant ATP 80NSSC18K1009.
KRA is supported by a Lyman Spitzer, Jr.~Postdoctoral Fellowship at Princeton University.

\bibliography{references.bib}
\end{document}